\documentstyle[preprint,prb,aps]{revtex}
\tightenlines
\begin{document}
\draft
\title{Wave-vector dependence of spin and density
multipole excitations in quantum dots}
\author{Manuel Barranco$^1$\cite{perm}, Leonardo Colletti$^1$,
Agust\'{\i} Emperador$^2$,
Enrico Lipparini$^1$, Mart\'{\i} Pi$^1$, and Lloren\c{c} Serra$^{3}$.}
\address{$^1$Dipartimento di Fisica, Universit\`a di Trento,
and INFM sezione di Trento, I-38050 Povo, Italy}
\address{$^2$Departament ECM,
Facultat de F\'{\i}sica,
Universitat de Barcelona, E-08028 Barcelona, Spain}
\address{$^3$Departament de F\'{\i}sica,
Universitat de les Illes Balears, E-07071 Palma de Mallorca, Spain}
\date{\today}

\maketitle

\begin{abstract}

We have employed  time-dependent local-spin density functional
theory to analyze the multipole spin and charge density
excitations in  GaAs-AlGaAs quantum dots.
The on-plane transferred momentum degree of freedom has been
taken into account, and the wave-vector dependence of the excitations
is discussed. In agreement with previous experiments,
we have found that the energies of these  modes do not depend
on the transferred wave-vector, although their intensities do.
Comparison with a recent resonant Raman scattering
experiment [C. Sch\"uller et al, Phys. Rev. Lett {\bf 80}, 2673 (1998)]
is made. This allows to identify 
the angular momentum of several of the observed modes as well as to 
reproduce their energies.
\end{abstract}

\narrowtext
\section*{}
\section{Introduction}

The characteristic single particle and collective excitations of typical
quantum dots (QD) are known to lie in the far-infrared (FIR) energy
region, i.e., they have energies that, depending on the size of the dot,
span the range from a few tens of meV to a fraction of meV.
Experimental information about FIR spectra was first obtained from photon
absorption experiments on InSb and on GaAs quantum dots\cite{Sik89,Dem90}.
Since the confining potential for small dots is parabolic to a good
approximation, and in the FIR regime the dipole approximation works
well, the absorption spectrum is rather insensitive to the number
of electrons in the dot, measuring to a large extent only the
center-of-mass excitations, which at non zero magnetic fields ($B$)
correspond to the two allowed dipole transitions arising from
each of the two possible circular polarizations of the absorbed
light. Two limitations of the absorption process, namely that it is
dominated by the $L$ = 1 multipole of the incoming electromagnetic wave,
and its insensitivity to the electronic spin degree of freedom, have
motivated that  theorists have been mostly concerned with the study of
dipole charge density  excitations (CDE). Yet,
higher multipolarity CDE's have been discussed using a classical model
\cite{Shi91}, a Hartree-random phase approximation method
\cite{Gud91}, a classical hydrodynamical model \cite{Ye94}, and an
equation of motion method \cite{Lip97}. Quadrupole $L=2$ CDE's have
also been addressed for the quantum-dot helium\cite{Wag95}.

The situation is changing with the use of  inelastic light scattering
to  study QD excitations. This experimental technique is nowadays
recognized as one of the more powerful tools to study the elementary
excitations of low dimensional electronic
nanostructures\cite{Str94,Loc96,Sch96,Sch98,Gon91,Sch94}, and it is
contributing to a deeper understanding of the two dimensional
electron gas\cite{Eri99,Pel97,Pin91,Pin89,Abs84} (2dEG).
Using  polarization selection rules, it allows to disentangle CDE
from  spin density (SDE) and single particle excitations (SPE), and
to observe them all in the same sample. Moreover, it  offers the
 possibility of studying the wave-vector dispersion dependence of
the excitations.

Measurements of Raman scattering on high-quality GaAs-AlGaAs quantum
dots have been  reported\cite{Sch98}. What make this
experiment especially appealing is that sharp  spin and charge
density excitations have been measured in conventional backwards
geometry as a function of the applied magnetic field $B$  and
of the  transferred lateral wave-vector $q$. Previous studies
were carried out at zero magnetic field \cite{Str94,Sch96}, or
the experimental conditions were such that the spectra did not
show a wave-vector conservation nor a clear polarization dependence
\cite{Loc96}, hence it was not possible to resolve SDE and CDE from
SPE, nor to record the spectra at predetermined $q$ values. We attempt
here a theoretical interpretation of these results based on the
time-dependent local-spin density functional theory (TDLSDFT),
addressing  the description of high
multipolarity spin and charge density modes of a QD, and incorporating
in a realistic way the on-plane wave-vector dependence of these
collective excitations.

\section{TDLSDFT description of collective modes in QD}

The dipole longitudinal response of  quantum dots has been recently
addressed in detail \cite{Ser98,Ser99}. We sketch here
how the method can be generalized to deal with other multipolarities
and the wave-vector degree of freedom.

The first task is to obtain the ground state (gs) of the dot
 solving the appropriate Kohn-Sham (KS) equations. The
exchange-correlation
energy density ${\cal E}_{xc}(n, m)$, where $n$ is the electron
density and $m$ the spin magnetization, constitutes a  key
ingredient of the method. It has been  constructed from
the results of Ref. \onlinecite{Tan89} on the nonpolarized and fully
polarized 2dEG using the two dimensional von Barth and Hedin
prescription to interpolate between both regimes\cite{Bar72}.

Once the KS gs has been worked out, we have determined the
induced densities originated by an external excitation field
employing linear-response theory.
For independent electrons in the KS mean field,
the variation $\delta n^{(0)}_{\sigma}$ induced in the spin
density $n_{\sigma}$  ($\sigma\equiv\uparrow,\downarrow$) by an external
spin-dependent field $F$, whose non-temporal dependence we denote as
$F=\sum_{\sigma}f_{\sigma}(\vec{r})\,|\sigma\rangle\langle\sigma|$,
can be written as \cite{Wil83}
\begin{equation}
\delta n^{(0)}_{\sigma}(\vec{r},\omega) =
\sum_{\sigma'}\int d\vec{r}\,'\chi^{(0)}_{\sigma\sigma'}
(\vec{r},\vec{r}\,';\omega)f_{\sigma'}(\vec{r}\,')\; ,
\label{eq1}
\end{equation}
where
$\chi^{(0)}_{\sigma\sigma'}$ is the KS spin density correlation function.
In this limit, the frequency $\omega$ corresponds to the
harmonic time dependence of the external field $F$ and  of
the induced $\delta n^{(0)}_\sigma$. Eq.\ (\ref{eq1}) is a 2$\times$2
matrix equation in the two-component Pauli space.
In longitudinal response theory, $F$ is diagonal in this space,
and its diagonal components are written as a vector
$F\equiv\left(
\begin{array}{c} f_\uparrow\\ f_\downarrow\end{array}
\right) $.
We consider first the external $L$-pole fields
\begin{equation}
F^{(n)}_{\pm L}=  r^L e^{ \pm i L \theta}
\left(\begin{array}{c} 1\\ 1\end{array}\right)
\,\,\, {\rm and} \,\,\,
F^{(m)}_{\pm L}   =    r^L e^{\pm i L \theta}
\left(\begin{array}{c} 1\\ -1\end{array}\right)
\label{eq2}
\end{equation}
which cause, respectively, the charge and spin density $L$ modes.
For the monopole $L$ = 0 mode, these fields are simply taken proportional
to $r^2$ (see below). To distinguish the induced densities in each
excitation channel
they will be labelled with an additional superscript as
$\delta n^{(0,n)}_\sigma$ or  $\delta n^{(0,m)}_\sigma$.

The TDLSDFT induced densities are obtained from the integral
equations
\begin{eqnarray}
\delta n^{(A)}_{\sigma}(\vec{r},\omega)
=
\delta n^{(0,A)}_{\sigma}(\vec{r},\omega)
+
\sum_{\sigma_1\sigma_2}\int d\vec{r}_1d\vec{r}_2\,
\chi^{(0)}_{\sigma\sigma_1}(\vec{r},\vec{r}_1;\omega)
K_{\sigma_1\sigma_2}(\vec{r}_1,\vec{r}_2)\,
\delta n^{(A)}_{\sigma_2}(\vec{r}_2,\omega) \,\,\, ,
\label{eq3}
\end{eqnarray}
where either $A=n$ or $A=m$, and
the kernel $K_{\sigma\sigma'}(\vec{r},\vec{r}\,')$
is the electron-hole interaction.

Equations (\ref{eq3}) have been solved as a generalized matrix
equation in coordinate space. Taking into account
angular decompositions of  $\chi_{\sigma\sigma'}$ and
$K_{\sigma\sigma'}$ of the kind
$K_{\sigma\sigma'}(\vec{r},\vec{r}\,')=
\sum_{\ell}K^{(\ell)}_{\sigma\sigma'}(r,r') e^{i \ell(\theta -\theta')}
$, it is enough to solve them for each multipole separately
because only modes with $\ell=\pm L$ couple to the external $L$-pole
field. One has
\begin{eqnarray}
K^{(\ell)}_{\sigma\sigma'}(r,r') &=&
{2\over\pi^{3/2}}
{\Gamma(|\ell|+1/2)\over\Gamma(|\ell|+1)}
{r_<^{|\ell|}\over r_>^{|\ell|+1}}
K_{|\ell|}\left({r_<\over r_>}\right) +
\left.{\partial^2{\cal E}_{xc}(n, m)
\over\partial n_{\sigma}\partial n_{\sigma'}}
\right\vert_{gs}{\delta(r-r')\over 2\pi r}\; ,
\label{eq4}
\end{eqnarray}
where $K_n(x)$ is given by the hypergeometric function\cite{Gra80}
${\pi\over2}F(1/2, n+1/2; n+1; x^2)$, and $r_> (r_<)$ is the greater
(smaller) of $r,r'$.

For a polarized system having a non zero magnetization in the gs,
the $\pm L$ modes are not
degenerate and give rise to two excitation branches with
$\Delta L_z=\pm L$, where $L_z$ is the gs orbital
angular momentum.
The induced  charge or magnetization densities corresponding to
density and spin responses are given by
$\delta n^{(A)}=\delta n^{(A)}_\uparrow+
\delta n^{(A)}_\downarrow$ and
$\delta{m}^{(A)}=\delta n^{(A)}_\uparrow-
\delta n^{(A)}_\downarrow$.
From them, the dynamical polarizabilities in the density and spin
channels are respectively given by
\begin{eqnarray}
\alpha_{nn}(L,\omega)&=&
\int{dr r^{L+1} \delta n^{(n)}(r)}\nonumber\\
\alpha_{mm}(L,\omega)&=&
\int{dr r^{L+1} \delta{m}^{(m)}(r)} \; .
\label{eq5}
\end{eqnarray}
For each $L$ value, taking into account both $\pm L$ possibilities we
define $\alpha_{AA}^{(L)}(\omega) \equiv \alpha_{AA}(L,\omega)
+ \alpha_{AA}(-L,\omega)$. Their imaginary parts are proportional
to the strength functions  $S^{(L)}_{AA}(\omega)=
{\rm Im}[\alpha^{(L)}_{AA}(\omega)]/\pi$. The peaks appearing in the
strength functions are the CDE or SDE
excited by the external field. Analogously, the peaks appearing
in the strength function which results from using in the above equations
the KS  density variations $\delta n^{(0,A)}_{\sigma}$ instead of
the correlated ones $\delta n^{(A)}_{\sigma}$, correspond to the SPE.

An analysis based on the use of the above
multipole excitation operators
$r^L e^{\pm i L\theta}$ implies that no appreciable on-plane
momentum $\vec{q}$ is transferred to the system, i.e.,
 $q \approx 0$. This will become apparent below.
Even in this limit,  some interesting features
of the experimental spectra are reproduced. Moreover,  it
allows one to make contact with  FIR photo-absorption spectroscopy.
Yet,  a more  detailed analysis of Raman spectra calls for
introducing the $q$ dependence in a realistic way.
A first attempt has been
made in Ref. \onlinecite{Loc96}, although the analysis
of the measured Raman spectra was carried out using a
Hartree model which cannot address the spin degree of freedom on the
one hand, nor take into account the contribution of charge and
spin  density collective modes to the scattering cross
section on the other hand.

Hamilton and McWhorter\cite{Ham69} were the first in pointing
out the important
role played by spin density modes in the Raman scattering in GaAs.
Their original formulation has been further elaborated by Blum
\cite{Blu70}, and more recently
the inelastic charge and spin density scattering cross sections have
been discussed in terms of the charge $S_{nn}(q,\omega)$ and spin
 $S_{mm}(q,\omega)$ strength functions\cite{Kat85}
(often called dynamic structure functions)
\begin{eqnarray}
\frac{d^2\,\sigma^{C}}{d\omega_s\, d\Omega_s} &\propto& |\hat{\bf e}_i
\cdot \hat{\bf e}_s|^2 S_{nn}(q,\omega)
\nonumber
\\
& &
\label{eq6}
\\
\frac{d^2\,\sigma^{S}}{d\omega_s\, d\Omega_s} &\propto& |\hat{\bf e}_i
\times \hat{\bf e}_s|^2 S_{mm}(q,\omega) \,\,\, ,
\nonumber
\end{eqnarray}
where $\omega$ is the energy difference of the incoming and scattered
photon $\omega_i - \omega_s$, and $\hat{\bf e}_{i,s}$ are the
polarization vectors. We refer the reader to the review articles of
Refs. \onlinecite{Abs84,Kle75} for a thorough discussion.

The above expressions are deceptively simple, but this is somehow
misleading, as simplicity
arises from the approximations made to arrive at
them\cite{Blu70,Kat85}.
Yet, they are often used to describe resonant Raman scattering in
GaAs heterostructures\cite{Loc96,Kat85,Haw85,Luo93}.
These approximations might obscure the
comparison of the calculated modes with these detected by Raman
spectroscopy. We believe, however, that rather than testing 
the TDLSDFT description of charge and spin density excitations,
it may manifest the limitations of theoretical schemes based on Eq.
(\ref{eq6}) to analyze resonant Raman scattering. It is worth
to mention the application made by Wendler et al \cite{Wen96}
to resonant  Raman scattering in two electron quantum rings using
a more general expression for the cross sections, of hopeless
applicability to  the $N= 200$ quantum dot described in
Ref.\ \onlinecite{Sch98}.

To obtain $S_{nn}(q,\omega)$ and  $S_{mm}(q,\omega)$ within
TDLSDFT, instead of  considering the response to  multipole
operators, one has to  consider the plane wave operator
$e^{i \vec{q}\,\vec{r}}$  involved in the  inelastic scattering
process. It is convenient to expand it into Bessel
functions\cite{Gra80}:
\begin{eqnarray}
& &e^{i \vec{q}\,\vec{r}} =
\sum_{\forall L} i^L \,J_L(q r)\, e^{i L \theta} =
\nonumber
\\
& & J_0(qr) +
\sum_{L>0} i^L \,J_L(q r)\, (e^{i L \theta}+
 e^{-i L \theta}) \,\,\, .
\label{eq7}
\end{eqnarray}
Depending on the $q$ value, the number of terms in the expansion
may be large, but  the method is of direct applicability because
the different $L$ terms in the expansion do not couple. Physically, it
is also sound to make the expansion, since the
experimental results display quite distinct peaks whose multipolar
character can, in some cases, be identified  even at a transferred
momentum as large as $0.8 \times 10^5$ cm$^{-1}$ (see Fig. 2 of Ref.
\onlinecite{Sch98}). Moreover, in the small $q$ limit, the expansion of
the Bessel functions leads to the multipole excitation operators we have
previously considered. In particular, the $r^2$ operator used in the
monopole case arises from the first non trivial term in the expansion of
$J_0(qr)$. An $r^2$ term is also present in the quadrupole case, this
time multiplied by the angular operators $e^{\pm 2 i \theta}$.

The TDLSDFT response to the plane wave operator
can thus be obtained as in the multipole case substituting in Eq.
(\ref{eq2}) $r^L$ by $J_L(qr)$ and  $r^2$ by $J_0(qr)$, and keeping
as many terms in the expansion Eq. (\ref{eq7}) as needed. A criterion
to determine the number of terms to be  considered is provided by
the f-sum rule\cite{Ser99}. For a given $q$ value,  the f-sum rules
of the plane wave operator and of each $L$ component in Eq. (\ref{eq7})
read (in effective atomic units):
\begin{eqnarray}
m_1^{(nn)}[e^{i \vec{q}\cdot \vec{r}}] =
m_1^{(mm)}[e^{i \vec{q}\cdot \vec{r}}] &=& q^2 \frac{N}{2}\; ,
\nonumber\\
m_1^{(nn)}[J_L(qr) e^{i L \theta}] =
m_1^{(mm)}[J_L(qr) e^{i L \theta}] &=& 
\nonumber\\
\frac{1}{2} \int d \vec{r}\, n_0(\vec{r}\,) \left\{ 
\left[\frac{d J_L(qr)}{dr}\right]^2 \right.
&+& \left. \frac{L^2}{r^2} J^2_L(qr) \right\}
\,\,\, ,
\label{eq8}
\end{eqnarray}
where $n_0(\vec{r}\,)$ is the gs electron density.
The maximum $L$ value in the expansion  has been fixed so
as to fulfill the plane wave f-sum rule  within 95 \% or better.
As a further numerical test, the second Eq. (\ref{eq8}) has been used
 to check the accuracy in the calculation of the strength functions
multipole by multipole.

\section{Results}

As a case of study
we present  a theoretical interpretation of the results obtained in
Ref. \onlinecite{Sch98} for an $N$ = 200 electron quantum dot  of radius
$R$ = 120 nm in GaAs-AlGaAs. We have modeled the confining potential by
the Coulomb potential created by a positively charged jellium disk
of the same radius\cite{Pi98}.
The only free parameter in the calculation is the number of positive
charges in the disk, which has been set to $N^+$ = 404 to reproduce
as many spin and density modes as possible at $B=0$, with a special
emphasis in the dipole SDE. We want to stress that this particular
jellium disk plays no other role that creating a
confining potential easy to generate and vary in a controlled way
by simply changing $N^+$. The question of whether the system  is
charged or not is misleading; after all, it could not be more
'charged' than any $N$ electron dot confined by a parabolic potential.
Image charges representing the gates and usually not considered in
QD structure calculations will eventually make neutral the whole
system\cite{Sto96}.
Discarding a  parabolic potential because of the large number of
electrons in the  dot, other confining potentials
\cite{Loc96,Bro90,Gud95} and fitting parameter strategies might
have been considered. However, a thorough testing of the confining
potential for such a large dot would imply  to obtain the charge and
spin responses at $B=0$ for several multipoles.
Obviously, this is a very demanding task.
A more elaborated search could have improved the
results we are going to discuss, which in some cases are not in
full agreement with experiment. Figure \ref{fig1} shows the
electron densities at $B=0$, 3 and 6 T.

The choice of the spin dipole mode at $B=0$ as the experimental
quantity to be better reproduced in the fit is motivated by the
emphasis we want to put in  the spin channel results, and because
for this mode two distinct branches with positive and negative $B$
dispersions are seen in the experiment. The dipole CDE at $B=0$
would have been a more conventional choice, but
unfortunately its experimental value has not been reported
\cite{Sch96,Sch98}. It is worth  to point out that even if
there seems to exist a common belief that CDE's are well understood,
for multipolarities different from the thoroughly studied dipole mode
this belief does not stem from having confronted so far theory
with real experiments. It is still an open question
how quantitative is the agreement between theory and experiment when
several CDE's have to be simultaneously described for the same QD.

The range of $B$ values investigated in
this work corresponds to filling factors larger than 3.
Consequently, the use of other density functional approaches such as
current density functional theory (CDFT) better suited at high
magnetic fields \cite{Pi98,Fer94,Ste98} can be avoided. For a
discussion of the difficulties one has to face to obtain the
longitudinal response within  time-dependent
CDFT, we refer the reader to Ref. \onlinecite{Lip99}.

\subsection{$q \approx 0$ results}

We first present the results obtained at $q \approx 0$. The interest
in studying this  limit lies in the  experimental
observation\cite{Str94,Sch96}, thoroughly discussed at $B=0$,
that in QD's the  energies of
the excited  modes do not depend on the transferred wave-vector $q$.
This is at variance with the situation in nanowires and in the 2dEG,
constituting a clear signature of the 'zero dimension' character of
QD's.
What changes with increasing $q$ is the total strength (see the first
Eq. (\ref{eq8})), and how it is distributed among the different peaks.
We shall discuss these matters in the next Subsection.

Figures \ref{fig2} to \ref{fig5} represent the spin and charge strength
functions for $L$ = 0 to 3. In the $L\neq 0$ cases we have
indicated with a $-(+)$ sign 
the excitations caused by the $+L(-L)$
component of the $F$ operators\cite{conv}  
in Eq.\ (\ref{eq2}). They correspond
to the two possible circular polarizations of the light absorbed or
emitted in the excitation or deexcitation process.
We have found that the spin peaks are rather fragmented, especially
in the monopole case.  However, they still are collective modes,
with energies redshifted from the single particle ones  due to the
attractive character of the exchange-correlation vertex corrections.

We would like to draw the attention to the  $-$ type, low energy
octupole SDE which is seen in Fig. \ref{fig5} to carry an
appreciable strength at $\omega \sim$  2.5 meV for
$B$ = 2 T. When a magnetic field is perpendicularly
applied to a QD, it is well known that  low  energy  modes
in the density channel are dipole edge CDE's arising from intraband
transitions, while bulk interband transitions lie at higher energy.
That may change with increasing $L$, and it is
easy to see that this is indeed the case for SDE's.
An inspection to  the KS single electron energies shown in
Fig. \ref{fig6} reveals that at high $L$'s, interband electron-hole
excitations are at lower energies than intraband ones.
Since the  electron-hole interaction is weak in the spin  channel
(only the exchange-correlation energy contributes to it),
we have found that at $B$ = 2 T the lowest energy octupole
SDE is a mode built from  interband electron-hole excitations.
Still, one might consider it as an edge mode, as its existence
is only possible because of the finite size of the system.
When $B$ increases further, the spin density edge mode has again a $+$
polarization. In the $L=3$, this happens at $B=3$T.
We have found that the low energy CDE is always a
$+$ type excitation, whereas the high energy CDE's are $-$ type
excitations arising from the corresponding component of $F^{(n)}$.

Figures \ref{fig7} and \ref{fig8} display the $B$ dispersion
of the more intense CDE's and SDE's, respectively.
The cyclotron frequency appears as a
peak in the calculated SPE (KS) dipole response, and we have not plotted
it in Fig. \ref{fig7}. The solid symbols represent the
experimental data\cite{Sch98}. We have connected with lines the more
intense peaks obtained in the calculation of the strength, which
displays some fragmentation, especially for high $L$ and $B$ values
(see also Ref. \onlinecite{Gud91}). We recall that only for a pure
parabolic confinement of frequency $\omega_0$
and for the $L=1$ mode in the dipole approximation, 
generalized Kohn's
theorem\cite{Koh61} ensures that  CDE's are distributed
according to the classical dispersion laws $\Omega \pm \omega_c/2$,
with $\Omega^2 = \omega_0^2 + \omega_c^2/4$ and $\omega_c$ being
the cyclotron frequency. We also recall that the
adiabatic TDLSDFT we are employing fulfills 
generalized Kohn's theorem\cite{Ser99}.

It can be seen from these figures that the experimental data are
only partly explained, as not all the experimental modes
are quantitatively described. In both spin and charge density
channels, TDLSDFT reproduces the weak $B$ dependence of the $L$ = 0
mode found in the experiment at small $B$ values. Our calculation
confirms the $L$ = 0, 1, and 2  multipolarity assigned  in the
experiment to the lower SDE's, but cannot identify the origin
of the higher SDE, whose signal is weak and broad, as mentioned
in Ref. \onlinecite{Sch98}. We will see in
the next Subsection that including finite momentum transfer, as
in actual experiments, does not greatly clarify the situation.  

At $B$ = 0, the energies of the $L > 0$ spin density excitations
follow the simple rule $E_L \sim L E_1$. We attribute this to the
weakness of the electron-hole interaction in the spin channel.
The prominent role played by the strong electron-hole interaction
in the charge density channel causes that rule to fail for CDE's.

As a general trend, the strength carried by the positive $B$
dispersion branch corresponding to the high $L$ spin density
excitations diminishes as $B$ increases. We have also found that
the spin strength becomes more fragmented with increasing $L$,
whereas bulk and edge magnetoplasmons associated with the
$\pm L$ excitations are better defined modes.

The positive $B$ dispersion branches of the  CDE's reveal a
complicated pattern at intermediate $B$ values, quite different from
the expected classical one holding up to $B \sim$ 2-3 T, but that
however fits a large set of the experimental modes.  The behavior of
these branches  has an
interesting quantal origin, namely the formation of well defined Landau
bands for magnetic fields larger than a critical value. Above it, the
more intense high energy collective peaks mostly arise from  transitions
between Landau bands whose index $M$ differs in one unit, $\Delta M=1$.
Since these bands are made of many
single electron states with  different $\ell$ values and energies
rather $\ell$ independent if $B$ is high enough \cite{Pi98},
this explains the
otherwise striking quasi $L$-degeneracy of the plasmon energies, only
broken by finite size effects and the $L$ dependence of the
electron-hole
interaction. Other modes with $\Delta M = 2$ build branches satellite
of those formed by the more intense $L$-peaks, and are clearly seen
in the calculation. Satellite branches of this kind appear even in the
dipole case \cite{Dem90,Lor96}, and are a clear signature of
nonparabolic confinement\cite{Gud91,Ye94,Ser99}. We will see
below how these branches emerge at high $q$ and $B$ values.

In contradistinction with the positive $B$ dispersion branches of the CDE's,
the negative $B$ dispersion ones do not manifest the quasi $L$-degeneracy.
While interband electron-hole excitations at the bulk of the dot are rather
$L$ independent as we have just mentioned, the negative $B$ dispersion
branches  are built from intraband electron-hole excitations at the
dot edge, and these are quite distinct for different $L$ values (see
Fig. \ref{fig6}, and Fig. 5 of Ref. \onlinecite{Pi98} for instance).

\subsection{Finite $q$ results}

The linear response to the multipole fields described before cannot
tell what is the relative intensity of the different  charge or spin
density excitations.
This limitation is circumvented  using the plane wave
operator for which  $S_{nn}(q,\omega)$ and
 $S_{mm}(q,\omega)$ display the charge or spin density excitations with
 nonarbitrary  relative intensity, allowing one to ascertain in each
channel which $L$ modes are more probably excited at given $B$ and $q$
values. This is clearly seen in Figs. \ref{fig9} to \ref{fig14}.

Figure \ref{fig9} shows the CDE's and SDE's at $B=0$ for selected $q$
values used in Refs. \onlinecite{Loc96} and \onlinecite{Sch98} 
(we shall give $q$ in $10^5$ cm$^{-1}$). 
Several interesting features show up in this
figure. We see that for small $q$ values the dipole mode
takes most of the strength, and that for the $q$ values
employed in Ref. \onlinecite{Sch98},  the
strength is exhausted by the modes with $L\leq 3$.
Another interesting observation, in full agreement with experiments,
is that the peaks have no appreciable wave-vector
dispersion\cite{Str94,Sch96}.

For a given $L$, Fig. \ref{fig9} also reveals the mechanism by which
the strength evolves with increasing $q$. Up to the $q$ values of
Ref. \onlinecite{Sch98} only the lowest energy peak of each multipolarity 
is sizably excited, and with increasing $q$ strength is transferred  
from dipole to quadrupole, monopole and octupole, successively.
For  larger $q$ values, as those employed in Ref. \onlinecite{Loc96},
higher energy peaks of each multipolarity get predominantly excited.
Conspicuous peaks corresponding to the {\em second} dipole and 
quadrupole modes, respectively, can be clearly seen at $q=5$ and 
$\omega\sim 16$ meV and $\sim 17$ meV.
The same happens at finite $B$ values, as it is shown in Fig. \ref{fig10}
for $B=1$T.

Figures \ref{fig11} and \ref{fig12} show the evolution with $B$
of the spectra corresponding to the largest $q$ value of 
Ref.\ \onlinecite{Sch98}.
Figures \ref{fig13} and \ref{fig14} show the same for 
the largest $q$ used in Ref.\ \onlinecite{Loc96}.
As anticipated, Fig.\ \ref{fig12} does not 
help identify the nature of the high energy 
SDE detected in the experiment. However, the results at higher $q$
(Fig.\ \ref{fig14}) show at low $B$ a broad distribution 
of the SDE strength centered around the energy of that experimental mode. 
One is tempted to speculate that the higher SDE seen in the experiment is 
just the envelope corresponding to our higher $q$ spectrum, 
which is centered around the 
second dipole SDE. In this sense, it is worth to point out that
our LSDA is essentially equivalent to a contact interaction in the spin 
channel, and thus it may underestimate finite momentum effects.
Of course, a similar effect could also contribute to the broad 
features observed in the higher CDE's.  
The $B=6$ T panels in Figs. \ref{fig11} and  \ref{fig13} show that
CDE's have a tendency to bundle, the energy spacing between bundles
roughly being $\omega_c$. We have already discussed this effect at
$q=0$.

Finally, we have used our results at $q=1.32$ to estimate the ratio
$r=(\omega_{SPE}-\omega_{SDE})/(\omega_{CDE}-\omega_{SPE})$
for the more intense peaks. This ratio is a quantitative
measure of the many-electron interactions in the dot\cite{Sch96}.
At $B=0$ we have obtained $r\sim 0.11$, in good agreement with the 
experimental value\cite{Sch96}. This ratio decreases with increasing $B$; 
we have found that $r\sim 0.08$ at $B=6$T.

\section{Summary}

In this work we have thoroughly discussed spin and charge density modes
of different multipolarity in GaAs-AlGaAs quantum dots, as well as their
wave-vector dependence. This has allowed us to make a detailed comparison
with experimental data obtained from resonant Raman scattering. In
particular, our calculations reproduce the experimental finding
that the excitation energies of the modes do not depend on the
transferred wave-vector, although their intensities do. The ratio
$(\omega_{SPE}-\omega_{SDE})/(\omega_{CDE}-\omega_{SPE})$
is also reproduced.

We have been able to  compare the energies of several
spin and density modes arising in the same dot. After fitting the
value of the spin density dipole mode at zero magnetic field, the
energies of the spin density modes up to $L=2$ haven been
quantitatively reproduced as a function of $B$.

The origin of the high energy
spin density mode at $q=1.32$ has not been elucidated by our calculations,
although our results for larger $q$'s predict a very broad distribution 
of strength centered around this experimental value.
The analysis of the strength
function at the experimental wave-vector $q$ seems to indicate that
no appreciable strength is carried by modes with $L>3$, and that
a broad structure consisting of peaks of different multipolarity
$L\leq 3$ and polarization
appears between 1 and 3 T (see the appropriate panels
in Fig. \ref{fig12}). 

The top panel of Fig. \ref{fig1} gives a hint about the
difficulty to properly describe high $L$ modes with rather simple
confining potential models. While for $L=0$ to 2 the excitation operator
is probing the bulk region and part of the edge of the dot, for higher
multipolarities it is only sensitive to its outermost edge structure
\cite{Lip97}. Obviously, for a large $N$ dot this region is
very much influenced by the actual structure of the confining potential,
and one should expect the larger disagreements between theory and
experiment to appear for these modes.

In the charge density channel, the agreement between theory and
experiment is more qualitative. 
At $B=0$ one of the measured CDE's is between our calculated $L=2$ 
and $L=3$ modes and, as in the spin density channel, only the $L=0$
mode is not appreciably dispersed with $B$.
Yet, we have given an interpretation, and a fair
quantitative description, of the peaks measured at
intermediate $B$ values which lie between the $\omega_c$ and $2\omega_c$
lines. As indicated, we have given more weight in the
fitting procedure to reproducing the SDE's.

Finally, we would like to point out that in spite of the difficulties
in interpreting  resonant Raman scattering in terms of
spin and density modes arising only from excitations of the conduction
band electrons, several  features of the spectra
 are well described within time-dependent local-spin density
functional theory. A more quantitative description of some aspects
of the experimental
spectra would require to take fully into account the underlying
structure of the system beyond the simple, idealized
semiconductor model currently used to describe quantum dots, and likely
a more realistic confining potential in the case of high multipolarities.

\section{acknowledgments}
This work has been performed under grants PB95-1249 and PB95-0492
from CICYT, Spain, and 1998SGR00011 from
Generalitat of Catalunya. A. E. and M. B. (Ref. PR1997-0174)
acknowledge support from the DGES (Spain).

\begin{figure}
\caption[]{Electron density of the $N=200$ dot (in units of $10^{11}$ cm$^{-2}$)
at $B=0$, 3 and 6 T. The dimensionless
horizontal scale  can be transformed into a more conventional one
recalling that $q=1.32 \times 10^5$ cm$^{-1}$. The value of $J_L(qr)$
for $L=0$ to 4 is also shown in the top panel for illustrative purposes.
}
\label{fig1}
\end{figure}
\begin{figure}
\caption{ Monopole strength function in arbitrary units as a function
of energy.
The thick solid line represents the charge density strength, the dashed
line the spin density strength, and the thin solid line the single
particle strength.
}
\label{fig2}
\end{figure}
\begin{figure}
\caption[]{Same as Fig. \ref{fig2} for the dipole mode. The signs
indicate the circular polarization of the more intense peaks, see
Eq. (\ref{eq2}).
}
\label{fig3}
\end{figure}
\begin{figure}
\caption[]{Same as Fig. \ref{fig3} for the quadrupole mode.
}
\label{fig4}
\end{figure}
\begin{figure}
\caption[]{Same as Fig. \ref{fig3} for the octupole mode.
}
\label{fig5}
\end{figure}
\begin{figure}
\caption{Single electron energies as a function of orbital angular
momentum for $B$ = 2 T.
Our choice of $B$ pointing towards
$+z$ favors that single particle states of negative angular momentum
and upwards spin be occupied. To avoid dealing with single particle 
angular momentum quantum numbers which are mostly negative, the angular 
dependence of the single particle wave functions is written as 
$e^{-i\ell\theta}$ and hence, $\ell$ represents the orbital angular 
momentum changed of sign. 
The horizontal line represents the electron
chemical
potential. Full, upright triangles correspond to $\sigma = \uparrow$
states,
and the empty, downright triangles to $\sigma = \downarrow$ states.
Interband and intraband transitions with $\Delta \ell =$ 2, 3 and 4 are
represented to illustrate the energy crossing discussed in the text.
}
\label{fig6}
\end{figure}
\begin{figure}
\caption[]{Energies of the more intense CDE's as a function of $B$.
The lines
connect the more intense peaks corresponding to a given multipole,
and the solid symbols represent the experimental data\cite{Sch98}.
}
\label{fig7}
\end{figure}
\begin{figure}
\caption[]{Same as Fig. \ref{fig7} for the more intense SDE's.
}
\label{fig8}
\end{figure}
\begin{figure}
\caption[]{$B=0$ charge (solid lines) and spin density (dashed
lines) strengths in arbitrary units for
$q=0.23$, 0.8, 1.32 and  $5 \times 10^5$ cm$^{-1}$.
The multipolarity of the main peaks is indicated.
}
\label{fig9}
\end{figure}
\begin{figure}
\caption[]{Same as Fig. \ref{fig9} for $B=1$T.
}
\label{fig10}
\end{figure}
\begin{figure}
\caption[]{Charge density strengths in arbitrary units
for $q=1.32 \times 10^5$ cm$^{-1}$ and different $B$ values.
The multipolarity and polarization of the main peaks is indicated.
}
\label{fig11}
\end{figure}
\begin{figure}
\caption[]{Same as Fig. \ref{fig11} for the spin density strength.
}
\label{fig12}
\end{figure}
\begin{figure}
\caption[]{Charge density strengths in arbitrary units
for $q=5\times 10^5$ cm$^{-1}$ and different $B$ values.
The multipolarity and polarization of the main peaks is indicated.
}
\label{fig13}
\end{figure}
\begin{figure}
\caption[]{Same as Fig. \ref{fig13} for the spin density strength.
}
\label{fig14}
\end{figure}
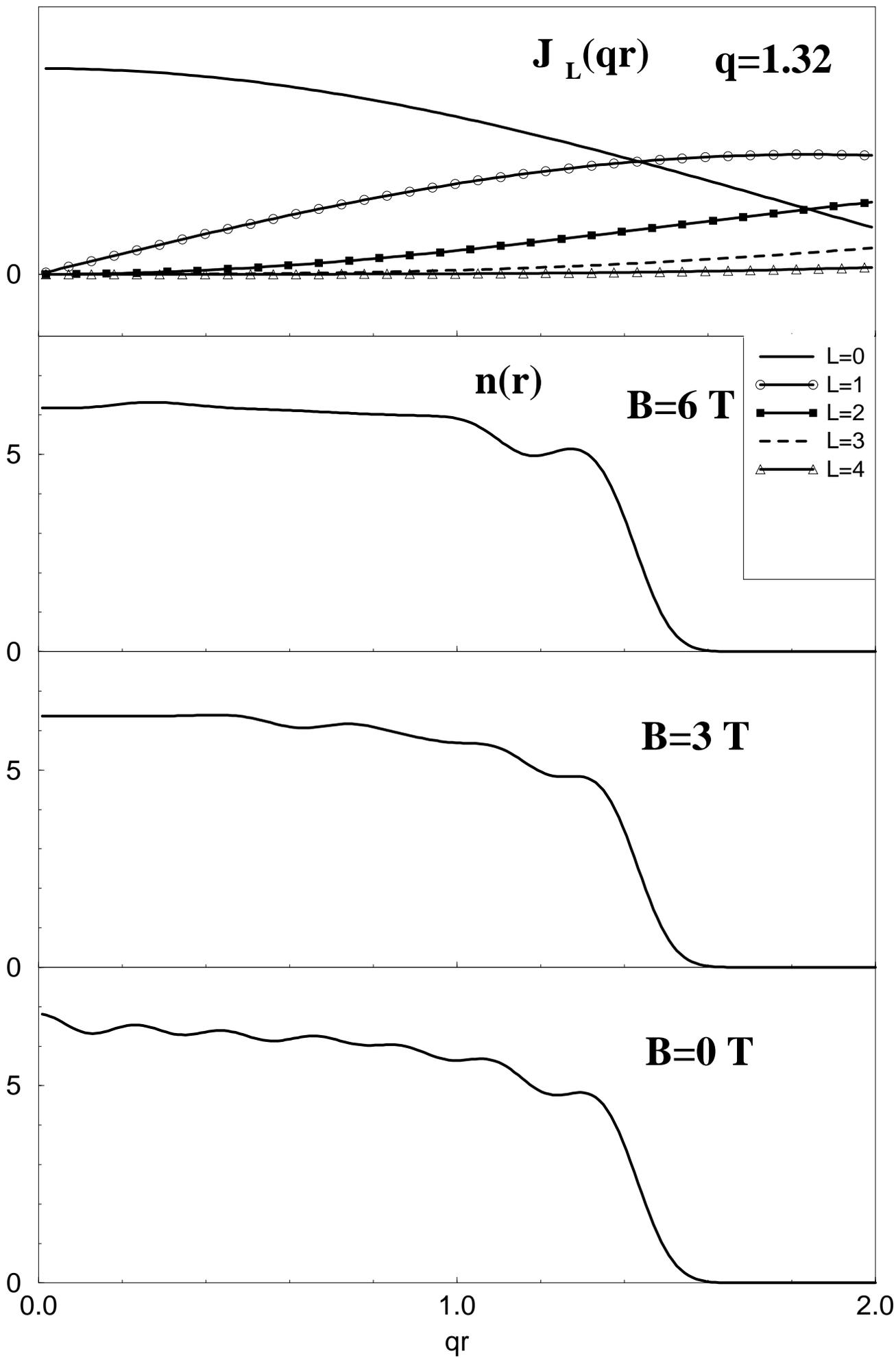
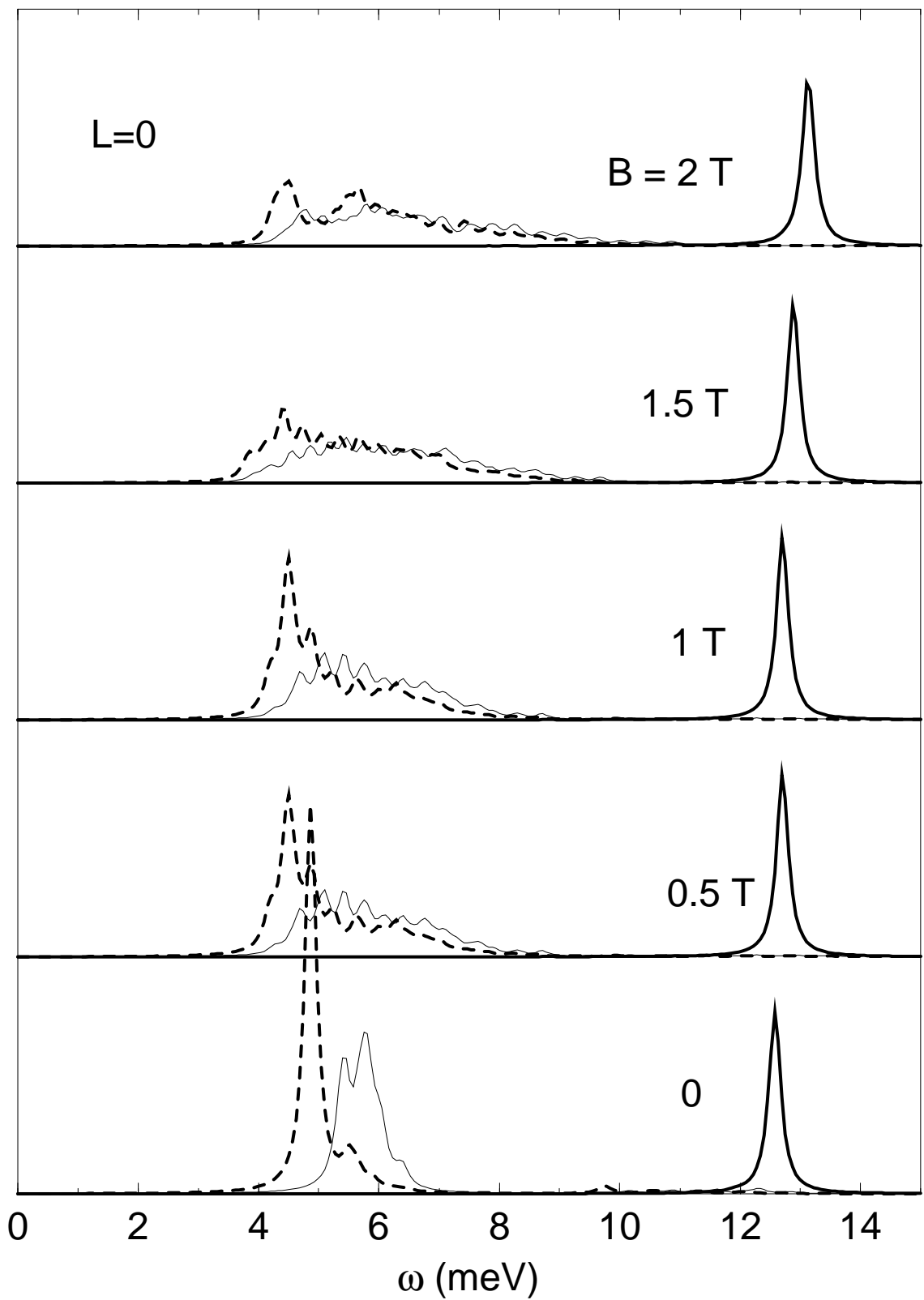
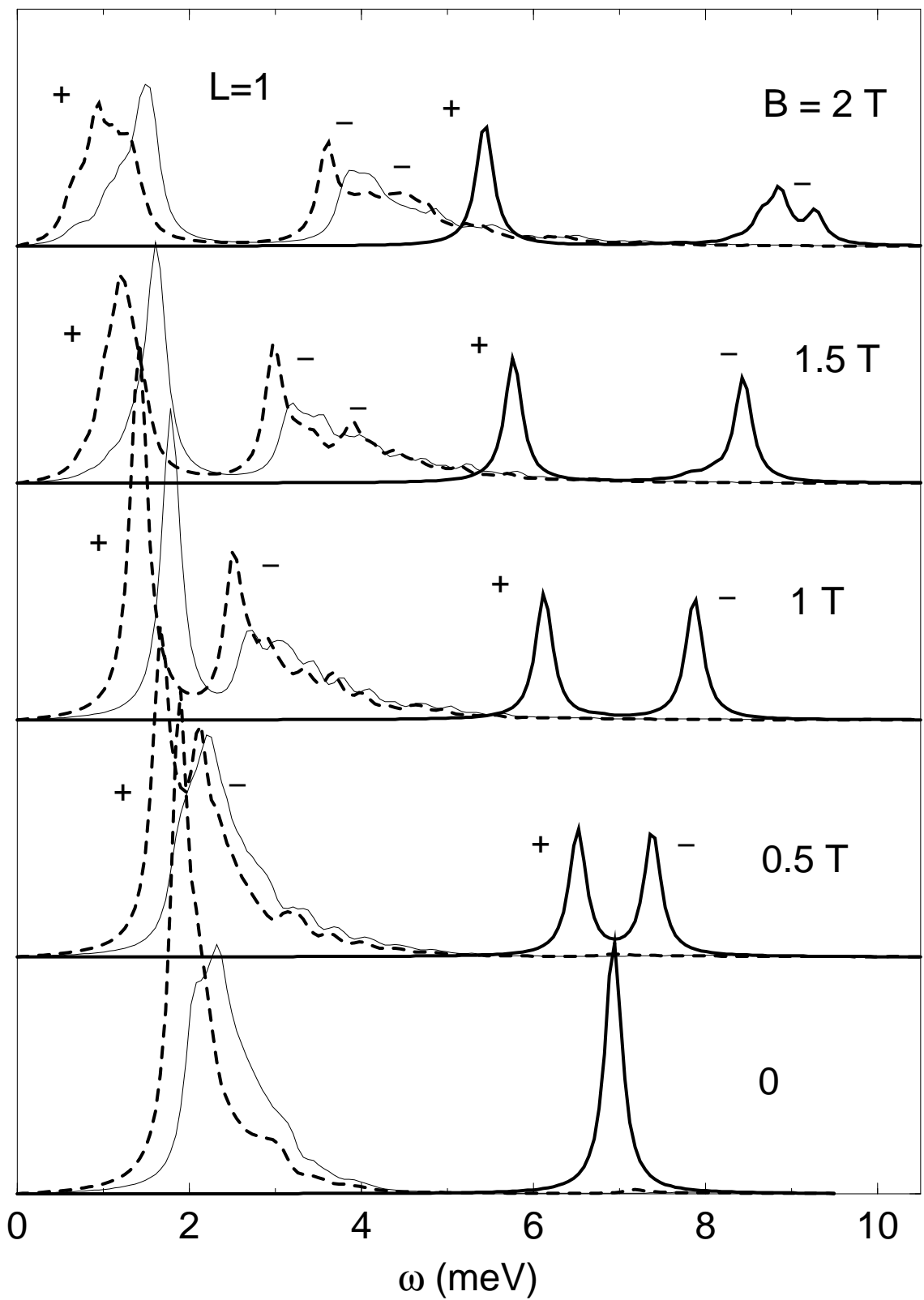
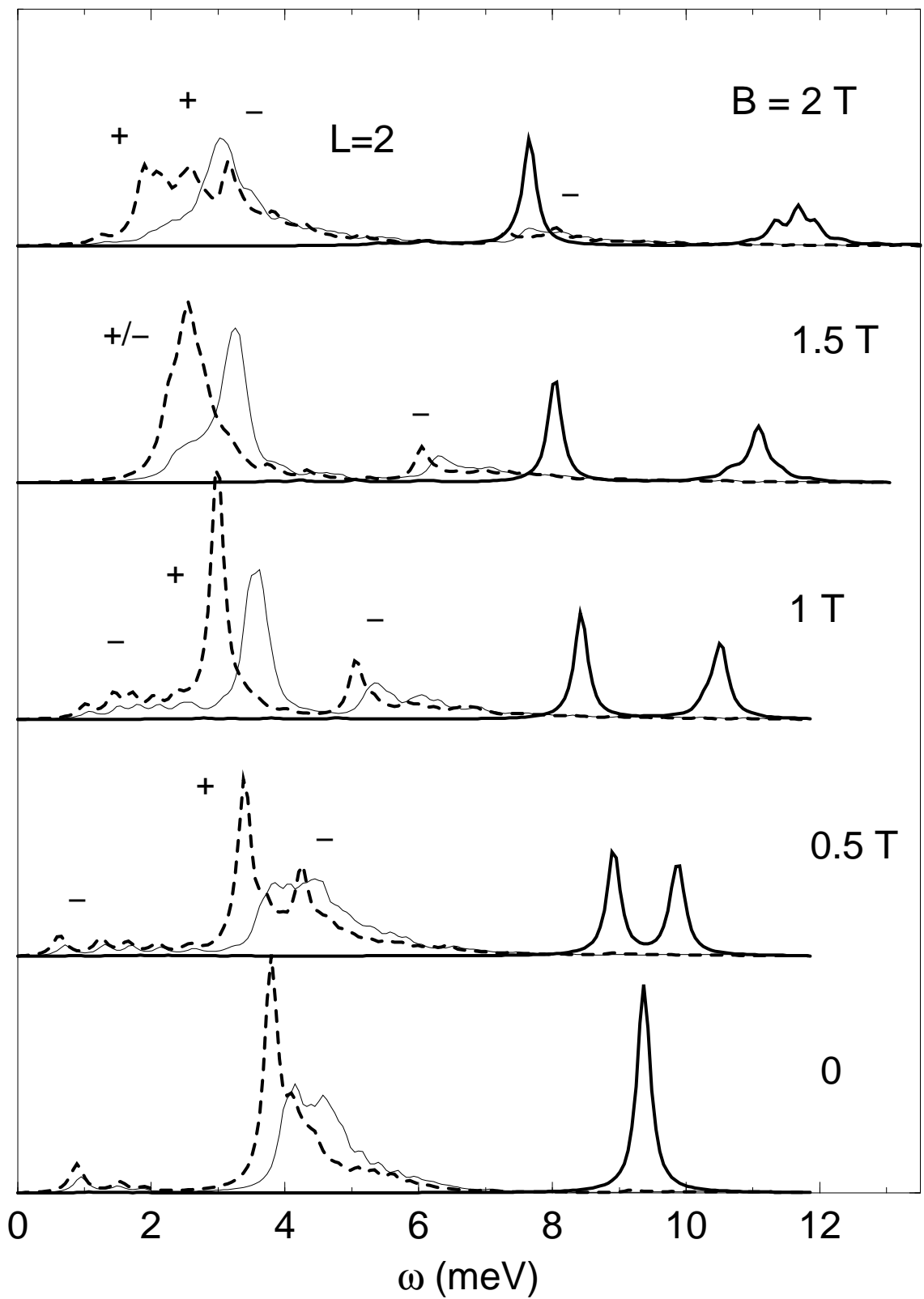
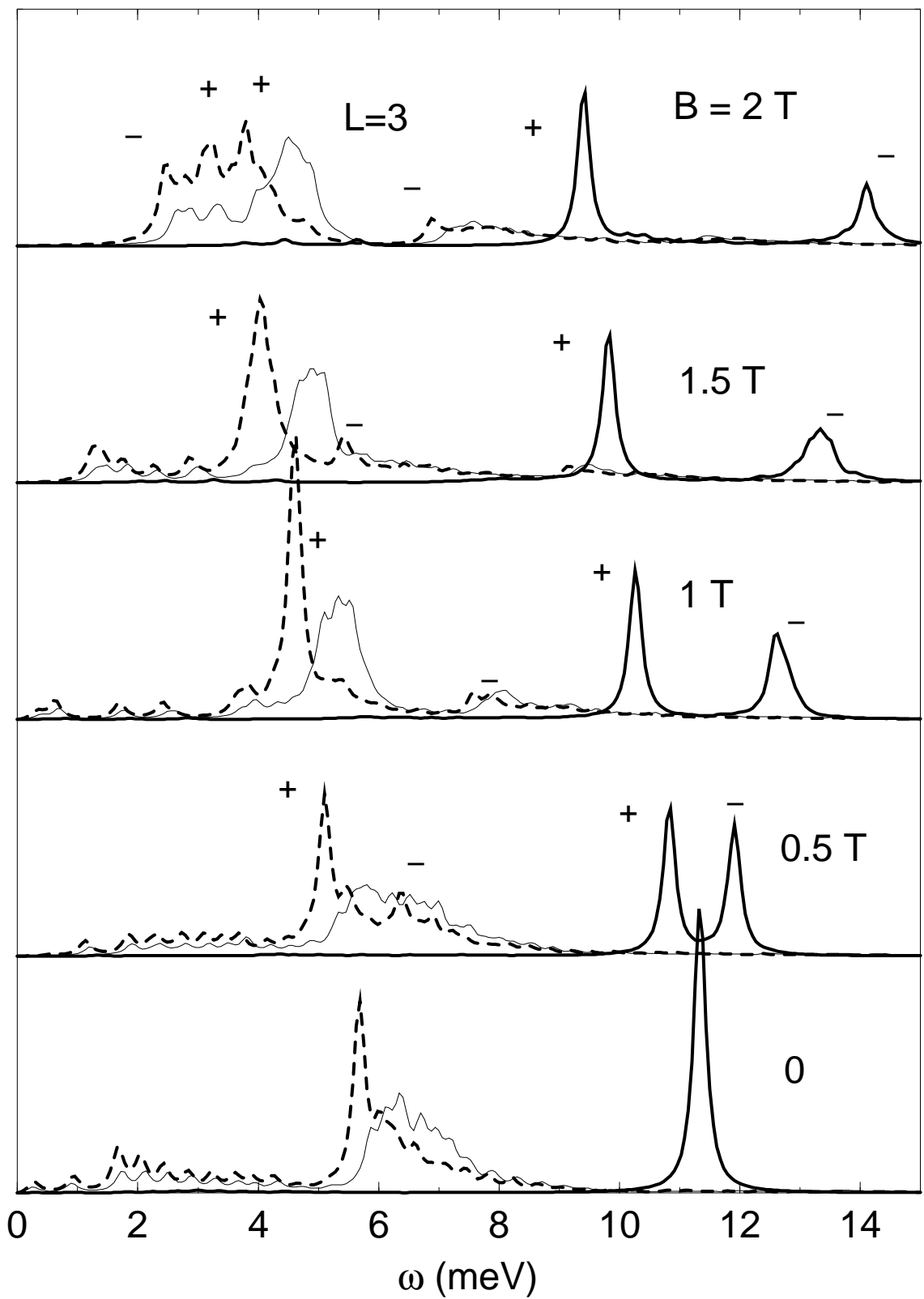
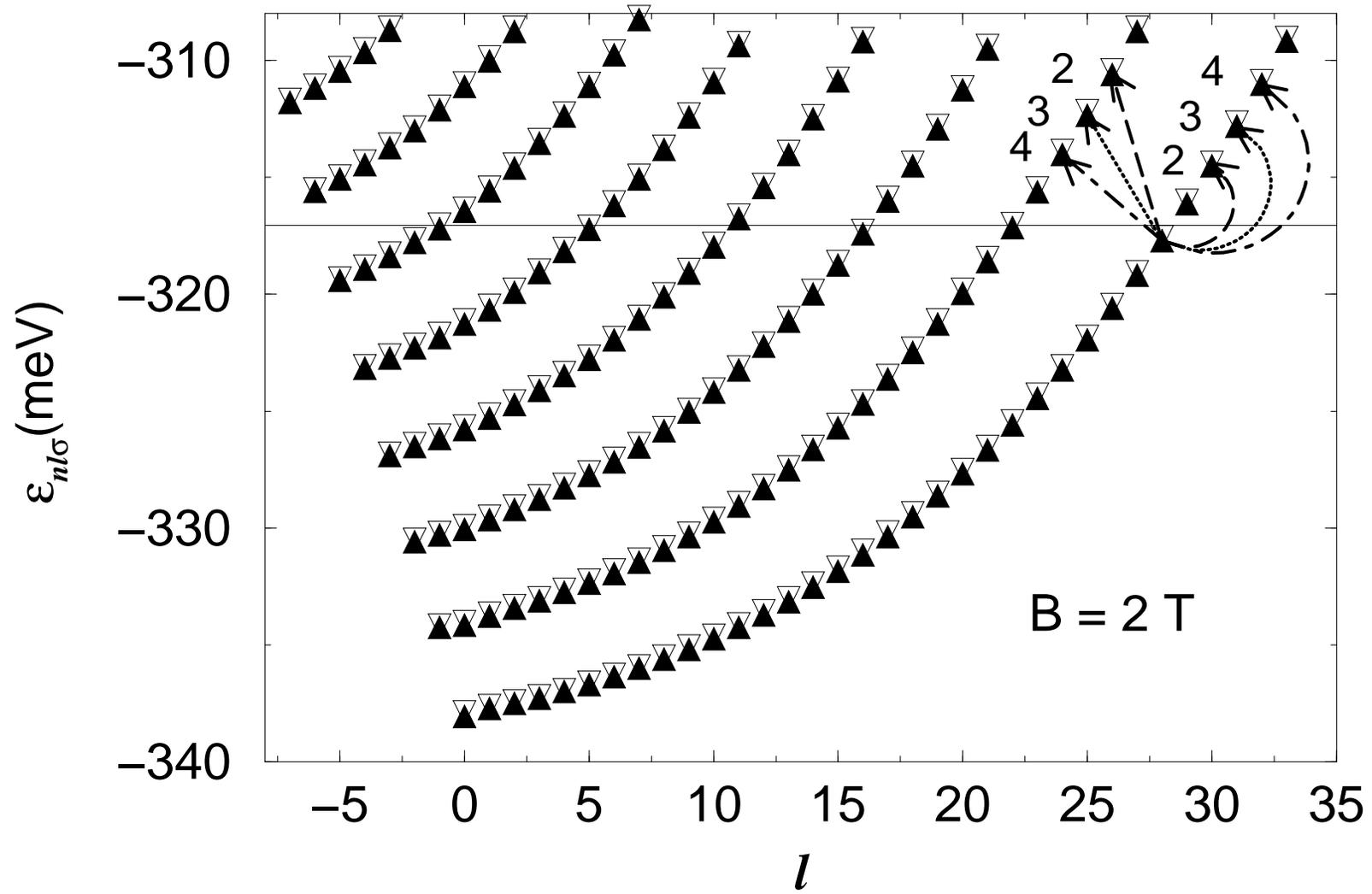
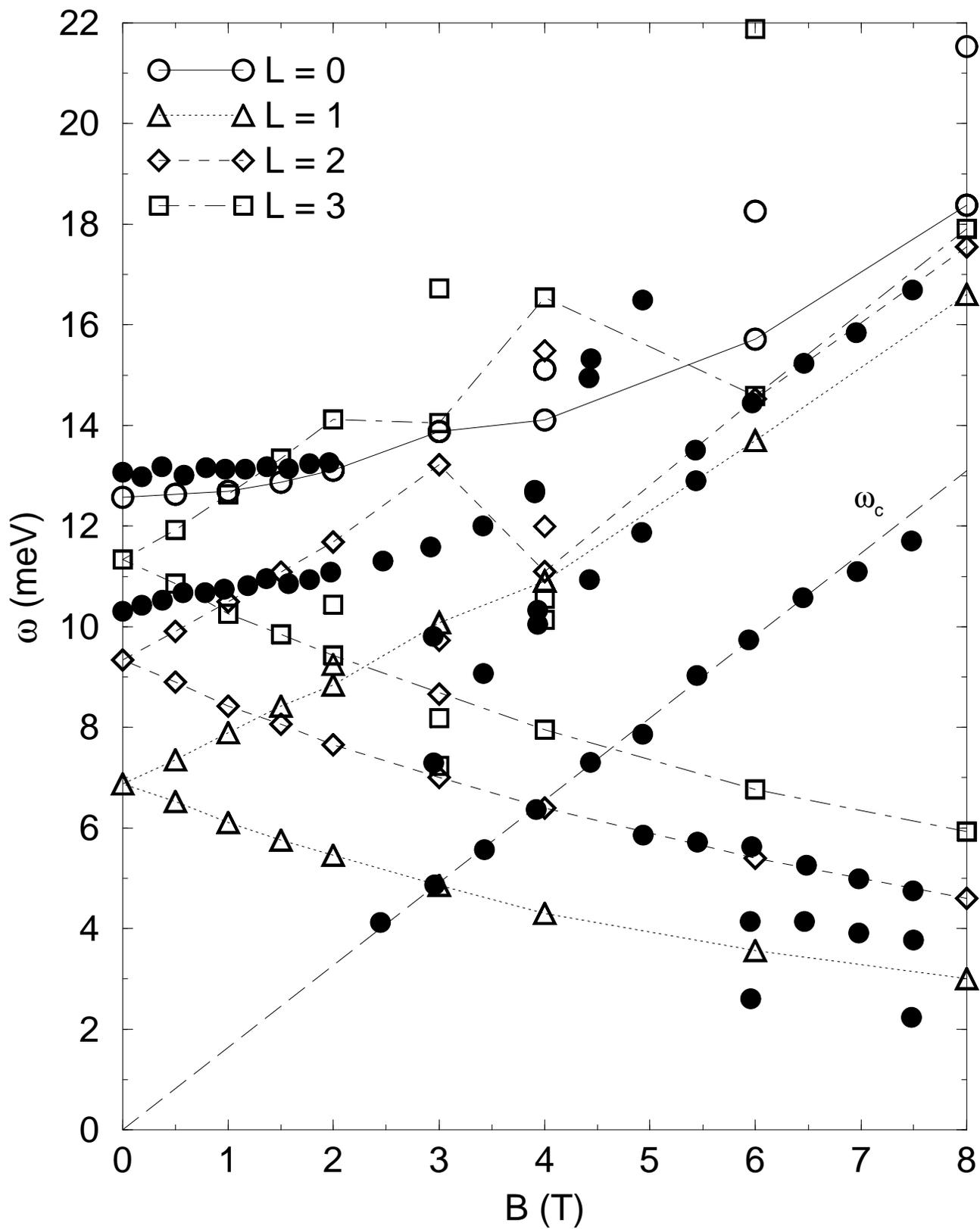
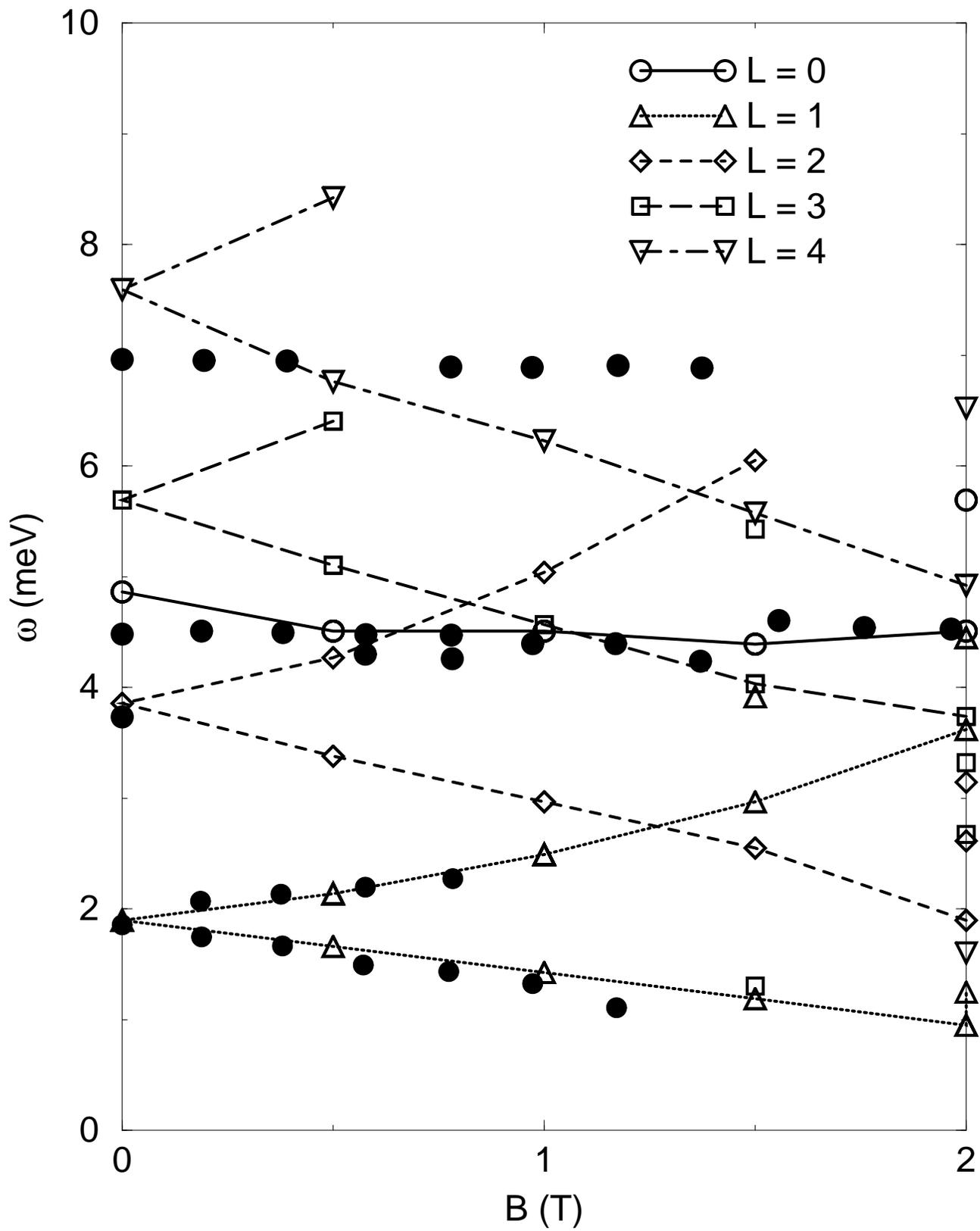
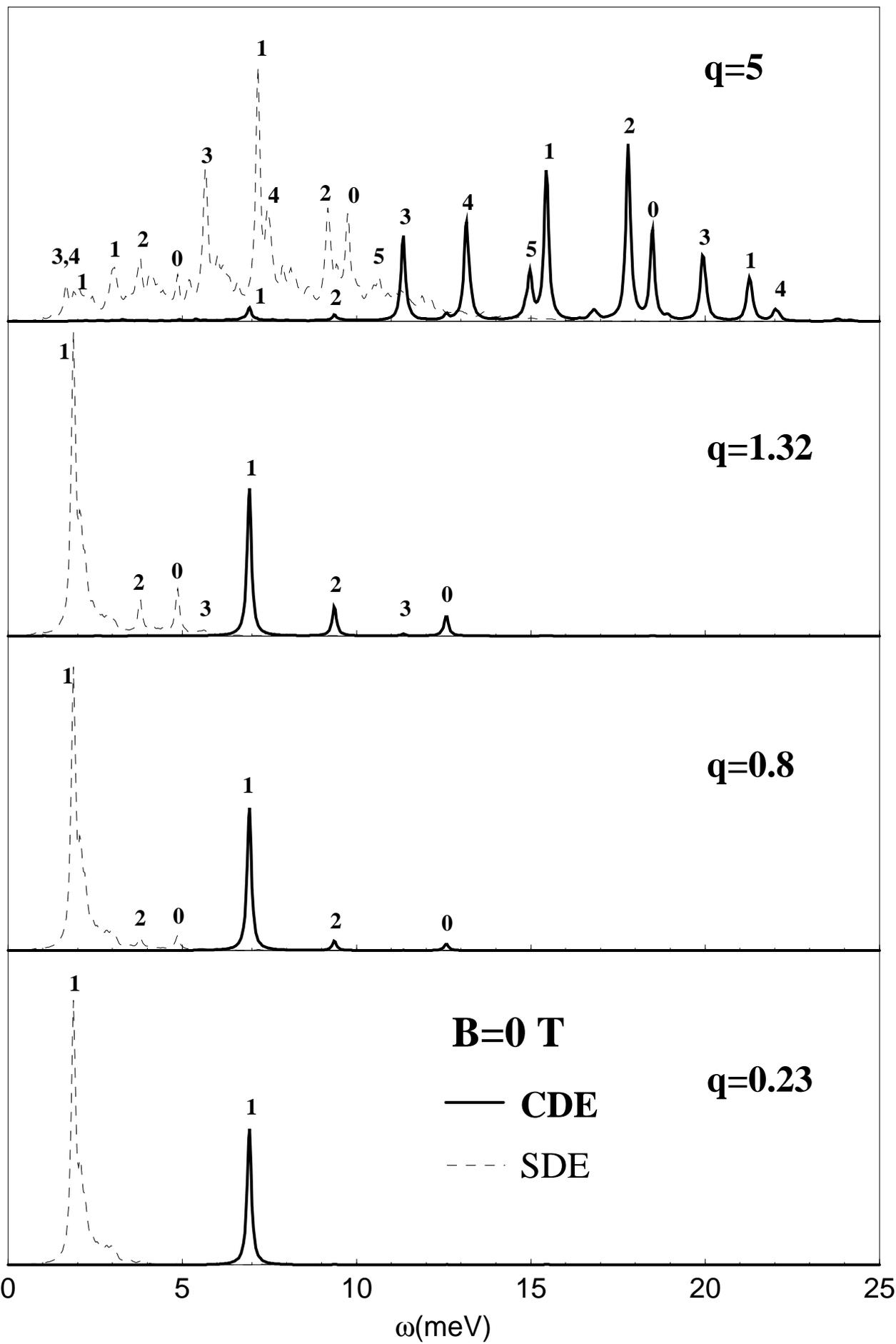
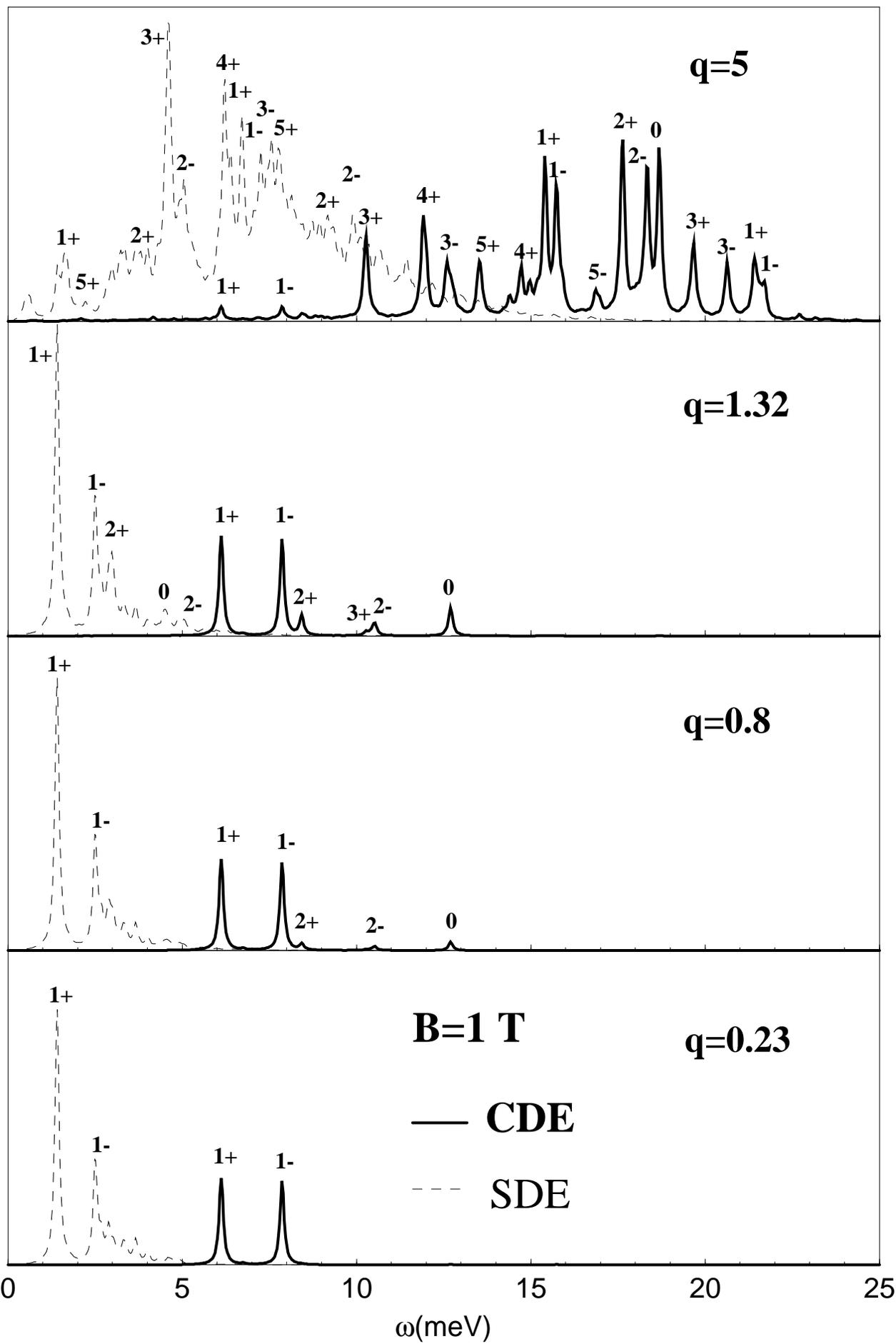
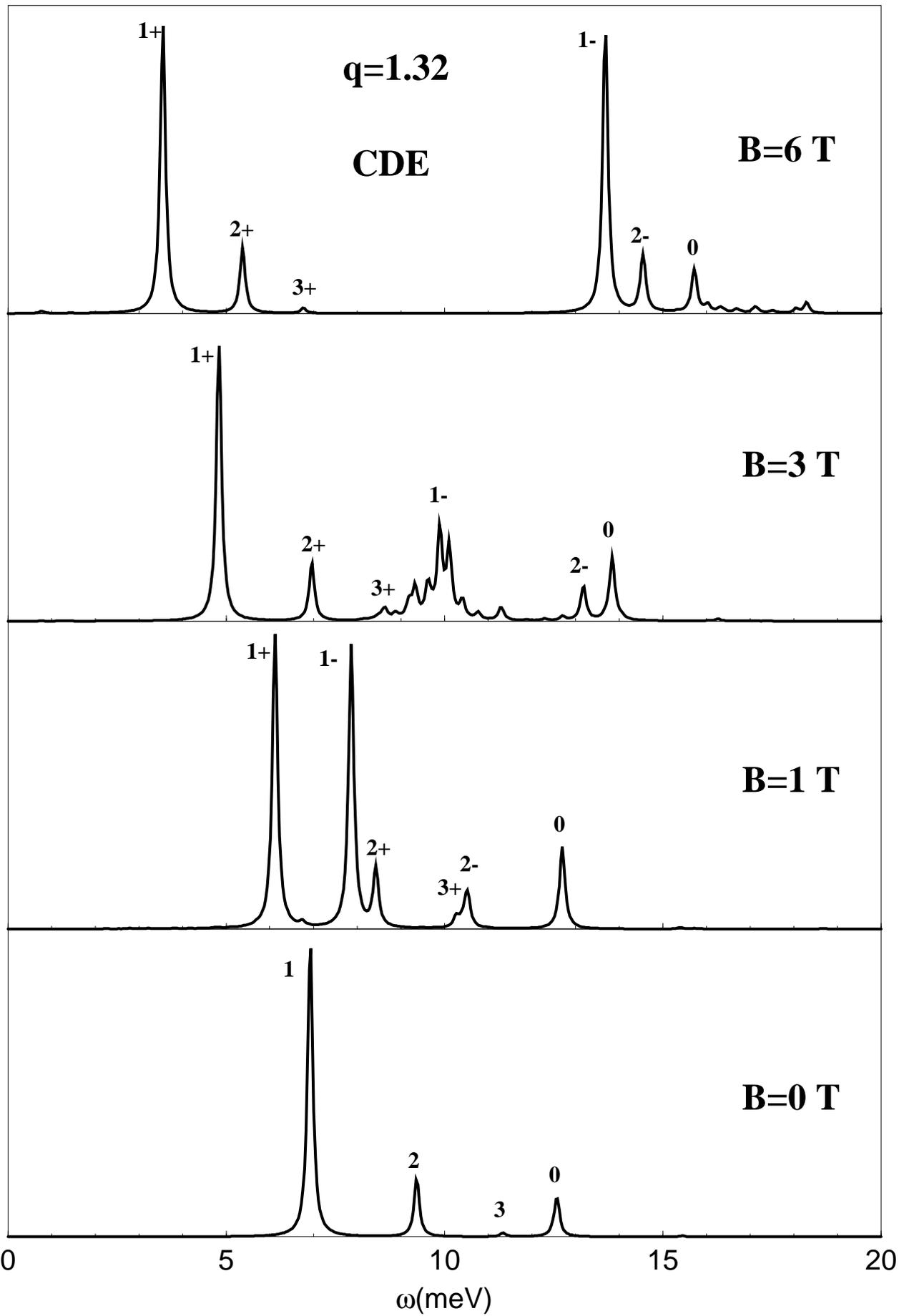
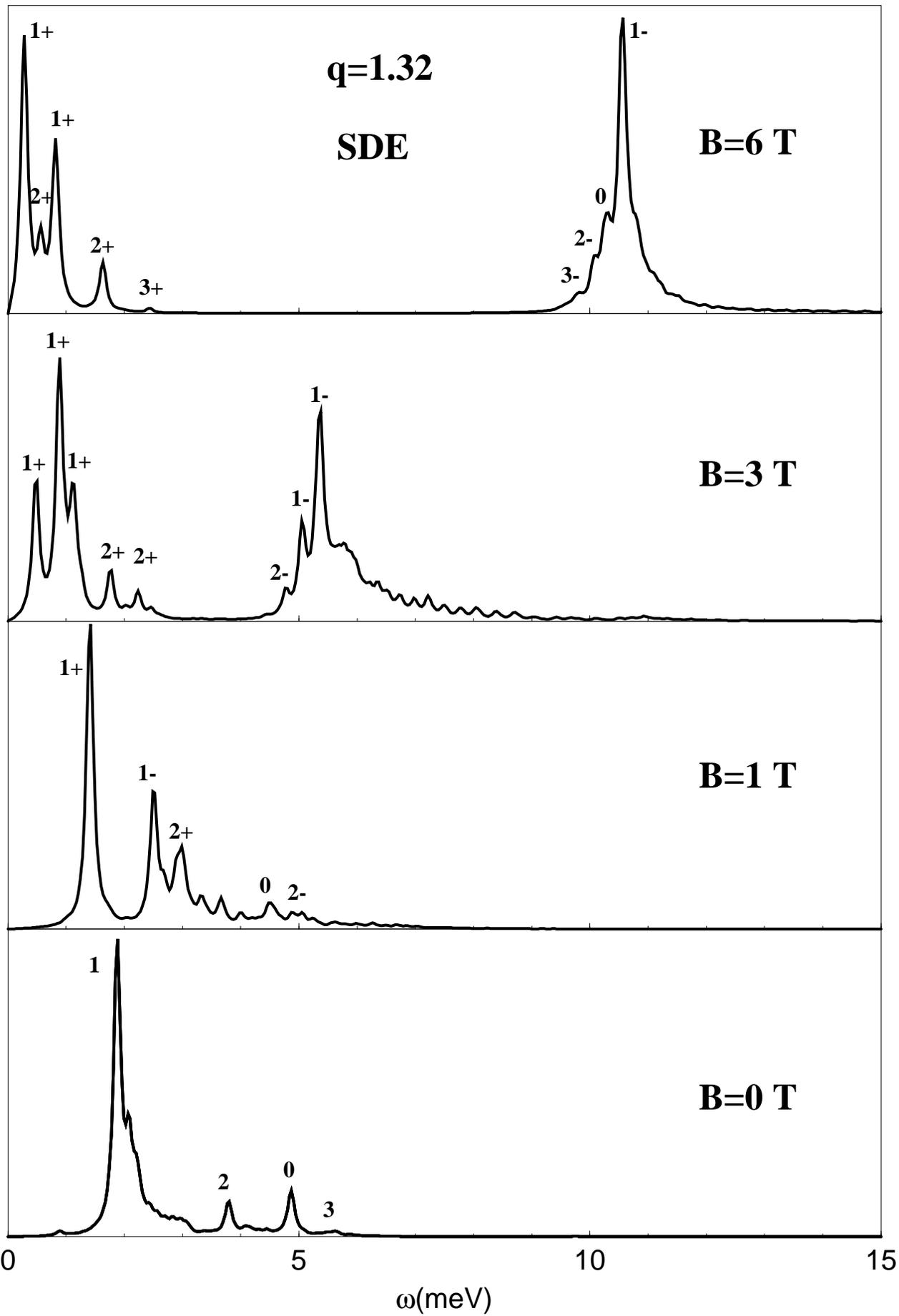
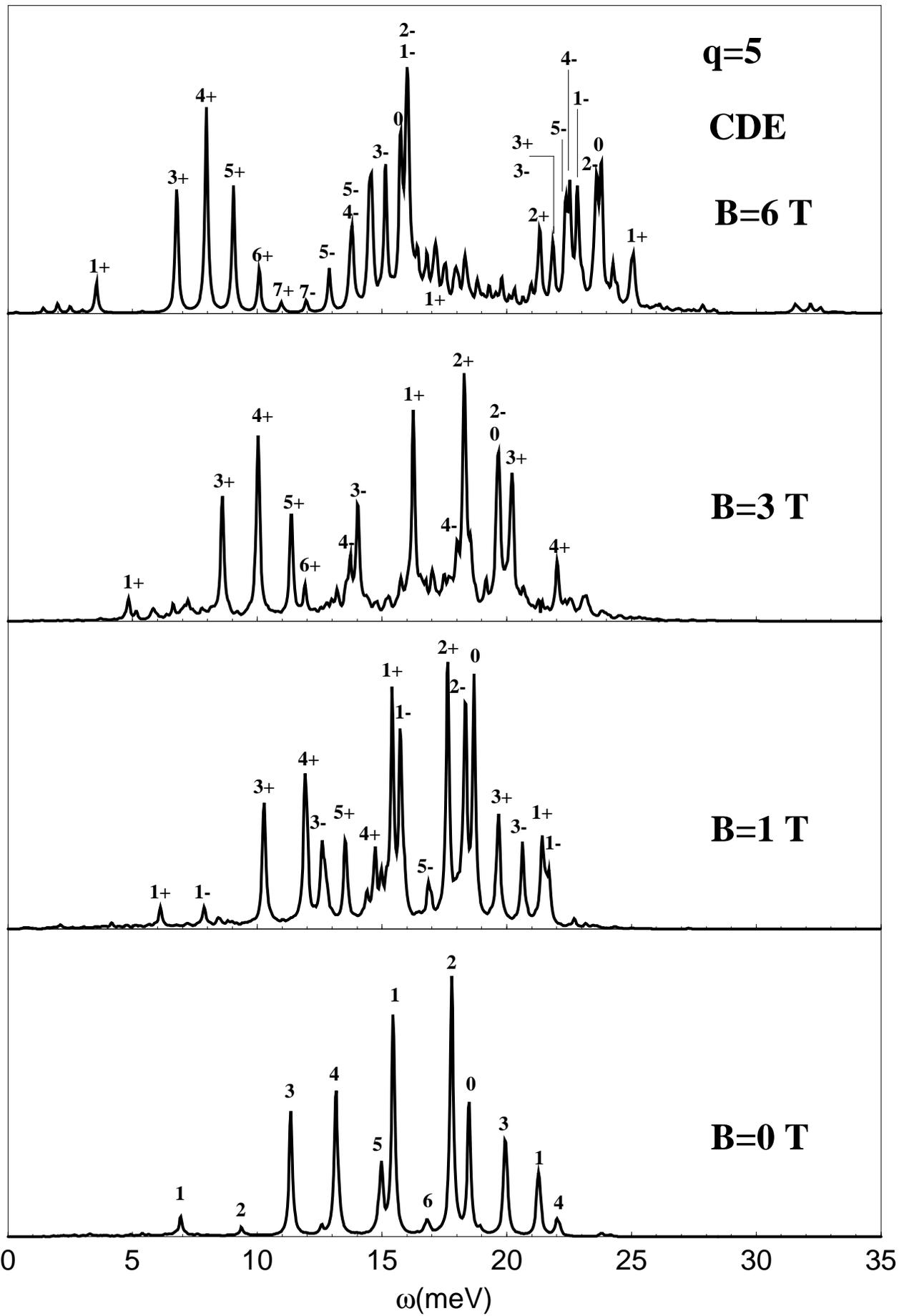
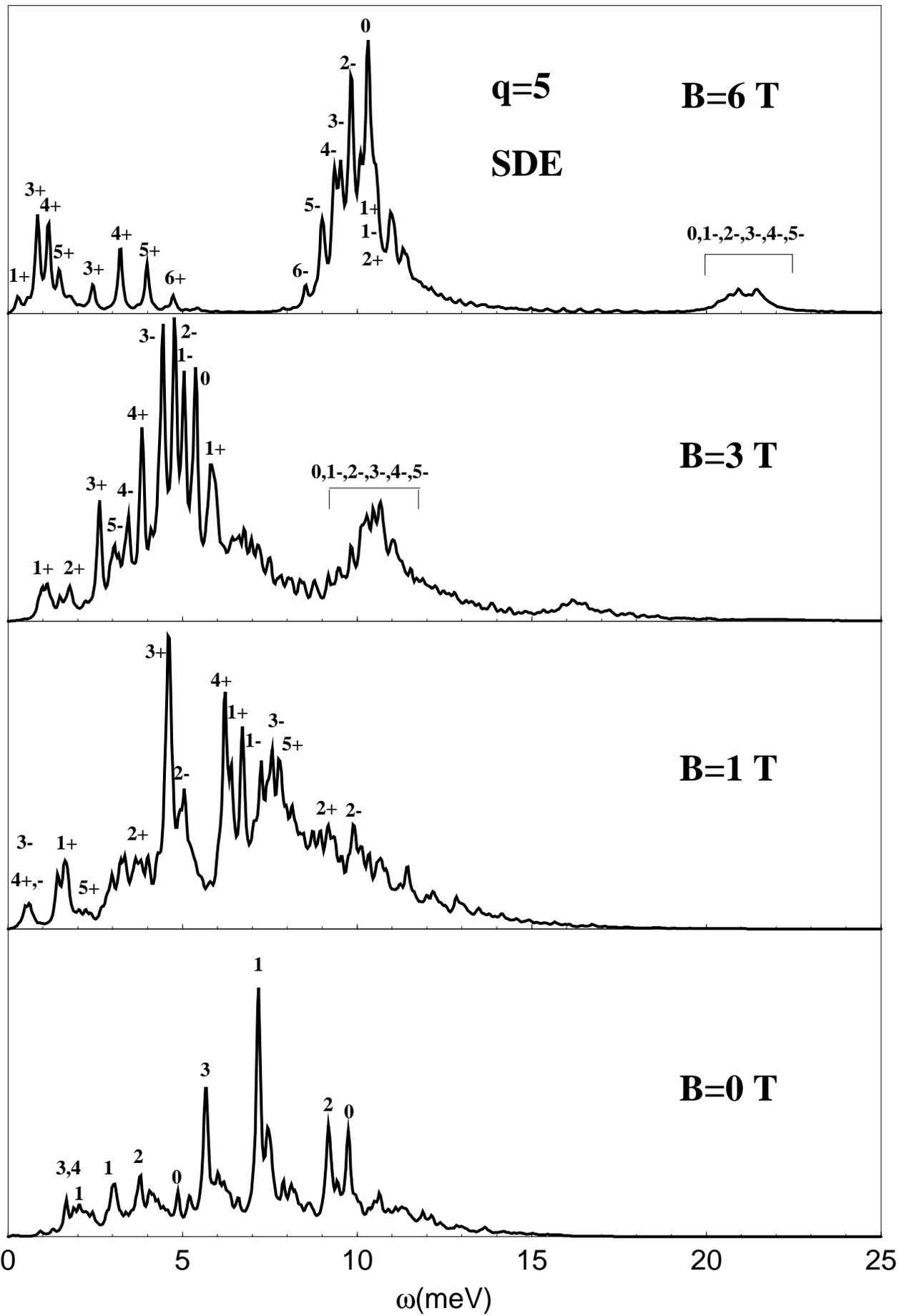
\end{document}